\documentclass[aps,prd,
nofootinbib,
preprintnumbers,
twocolumn,
showpacs, 
floatfix
]{revtex4}
\usepackage{amsmath}
\usepackage{slashed}
\usepackage{amssymb}
\usepackage{epsfig}
\usepackage{graphicx}
\usepackage{multirow}
\usepackage{color}
\usepackage[normal]{subfigure}
\usepackage{rotating}
\usepackage{hyperref}

\newcommand{\ewt}{\end{widetext}}
\newcommand{\be}{\begin{equation}}
\newcommand{\ee}{\end{equation}}
\newcommand{\bdm}{\begin{displaymath}}
\newcommand{\edm}{\end{displaymath}}
\newcommand{\bea}{\begin{eqnarray}}
\newcommand{\eea}{\end{eqnarray}}

\def\eq#1{{Eq.~(\ref{#1})}}
\def\eqs#1#2{{Eqs.~(\ref{#1})--(\ref{#2})}}
\def\fig#1{{Fig.~\ref{#1}}}

\def\Table#1{{Table~\ref{#1}}}

\def\sect#1{{Sect.~\ref{#1}}}

\def\app#1{{Appendix~\ref{#1}}}
\def\apps#1#2{{Apps.~\ref{#1}--\ref{#2}}}
\def\vev#1{\left\langle #1 \right\rangle}

\begin{document}
\preprint{TTP13-018, SFB/CPP-13-34, CP3-Origins-2013-017, DIAS-2013-17, SISSA 22/2013/FISI}
\title{A framework for baryonic R-parity violation in grand unified theories}

\author{Luca Di Luzio$^{a}$}
\author{Marco Nardecchia$^{b}$}
\author{Andrea Romanino$^{c}$}
\affiliation{$^{a}$Institut f\"{u}r Theoretische Teilchenphysik,
Karlsruhe Institute of Technology (KIT), D-76128 Karlsruhe, Germany}
\affiliation{$^{b}$CP3-Origins and Danish Institute for Advanced Study (DIAS), University of Southern Denmark, Campusvej 55, DK-5230, OdenseM, Denmark}
\affiliation{$^{c}$SISSA/ISAS and INFN, I--34136 Trieste, Italy}

\begin{abstract}
We investigate the possibility of obtaining sizeable R-parity breaking interactions violating baryon number but not lepton number within supersymmetric grand unified theories. Such a possibility allows to ameliorate the naturalness status of supersymmetry while maintaining successful gauge coupling unification, one of its main phenomenological motivations. We show that this can be achieved without fine-tuning or the need of large representations in simple SO(10) models. 
\end{abstract}

\maketitle

\section{Introduction}
\label{intro}

Supersymmetric scenarios without R-parity~\cite{Weinberg:1981wj,Sakai:1981pk,Aulakh:1982yn,Hall:1983id,Zwirner:1984is,Ross:1984yg} 
have received a renewed interest after the negative results of 
supersymmetry (SUSY) searches at the LHC. R-parity accounts for the stability of the lightest supersymmetric particle (LSP), whose escape from the detector gives rise to the prototypical supersymmetry signal: missing energy. R-parity violation (RPV) may allow  supersymmetric particles to evade the latter, stringent searches. 
In particular, it has been argued that scenarios in which R-parity is violated through 
baryon-number-violating interactions  could be particularly suited to hide supersymmetric 
signals into QCD backgrounds, thus implying a significant reduction of the current LHC 
lower bounds on the mass of the superpartners, hence the intense research 
activity on the subject in the recent years \cite{Brust:2011tb,Csaki:2011ge,Arcadi:2011ug,Allanach:2012vj,Dreiner:2012wm,Allanach:2012tc,KerenZur:2012fr,Dupuis:2012is,Brust:2012uf,Ruderman:2012jd,Evans:2012bf,Berger:2012mm,Asano:2012gj,Han:2012cu,Franceschini:2012za,Krnjaic:2012aj,Bhattacherjee:2013gr,Franceschini:2013ne,Csaki:2013we,Berger:2013sir,Florez:2013mxa,Krnjaic:2013eta,Monteux:2013mna}.

For baryon number violating RPV operators to be sizeable enough to hide supersymmetric particles, lepton number violating operators should be very suppressed, possibly absent. 
The simultaneous presence of $\Delta B \neq 0$ and $\Delta L \neq 0$ interactions is in fact extremely constrained by matter stability. Indeed, R-parity was originally introduced in order to obtain (accidental) lepton and baryon number conservation in the minimal supersymmetric standard model (MSSM), thus protecting it from renormalizable sources 
of a potentially way too large proton decay rate and neutrino masses. However, it is known that it suffices to assume the absence of R-parity lepton number violating operators, by means of a ``leptonic R-parity'', to get rid of such sources~\cite{Hall:1983id,Zwirner:1984is}. 

Introducing baryonic RPV  is therefore relatively safe if leptonic RPV is absent. On the other hand, one can wonder whether such an asymmetry between lepton and baryon number violating operators is compatible with grand unified theories (GUTs). After all, one of the motivations to persist on supersymmetric models despite the lack of signals is the very success of supersymmetric grand unification. This is the issue we would like to address in this paper. 

In the presence of grand unification, the natural expectation is that baryonic and leptonic RPV couplings are either absent or simultaneously present, as quarks and leptons share the same grand unified multiplets~\cite{Martin:1992mq,Lee:1994je,Mohapatra:1996fu,Mohapatra:1996iy,Aulakh:1999cd,Aulakh:2000sn}. Indeed, exact SU(5) invariance forces baryonic RPV to be accompanied by leptonic RPV. However, a source of asymmetry between the two types of RPV can be generated by SU(5) breaking. 

To be more specific, let us state our problem in the following terms: we would like to find a supersymmetric GUT for which the low-energy limit, well below the unification scale $M_G$, is described by the MSSM field content and gauge group and by a superpotential whose renormalizable part is given by
\begin{equation}
\label{eq:Weffren}
W_\text{ren} = W_\text{MSSM} + \lambda''_{ijk} u^c_i d^c_j d^c_k ,
\end{equation}
where 
$\lambda''_{ijk}$ is antisymmetric in the flavour indices $j, k$. 
The extra operator violates R-parity and baryon number ($\Delta B = -1$). Since grand unified gauge groups transform leptons into baryons (preserving $B-L$ in the minimal case of SU(5)), one would expect that operator to be accompanied by RPV and lepton-number violating ($\Delta L = 1$) operators such as $\lambda_{ijk} e^c_i l_j l_k$ and $\lambda'_{ijk} q_i d^c_j l_k$. Indeed, in minimal SU(5) grand unification $d_i^c$ and $l_i$ are unified in a $\overline{5}_i$ and $q_i$, $u^c_i$, $e^c_i$ are unified in a $10_i$ and the three above operators all come from $\Lambda_{ijk} 10_i  \overline{5}_j\overline{5}_k$, which gives $\lambda_{ijk} = \tfrac{1}{2} \lambda'_{ijk} = \lambda''_{ijk} = \Lambda_{ijk}$. In this case, the bounds from matter stability require $\Lambda_{ijk}$ to be smaller than at least $10^{-10}$ for any value of $i,j,k$  and for superpartners around the TeV scale \cite{Smirnov:1995ey}. Such tiny couplings would be irrelevant for collider physics since the LSP would be stable on the scale of the detector size. We then need to find a way to obtain sizeable $\lambda''$ couplings together with vanishing $\lambda,\lambda'$. 

While leptonic RPV in GUTs has been investigated in a number of papers, see e.g.~\cite{Hall:1983id,Brahm:1989iy,Smirnov:1995ey,Tamvakis:1996dk,Vissani:1995ha,Vissani:1996ch,Barbieri:1997zn,Giudice:1997wb,Arcadi:2011yw,Graham:2012th}, to our knowledge, such a problem was only considered in the context of SU(5) by Smirnov and Vissani~\cite{Smirnov:1995ey} and by Tamvakis~\cite{Tamvakis:1996np}.\footnote{There also exist models of baryonic R-parity violation in Flipped-SU(5) \cite{Tamvakis:1996np,Giudice:1997wb} and $\rm{SU(5)} \otimes \rm{SU(3)}$ \cite{Bhattacherjee:2013gr}.}  In~\cite{Smirnov:1995ey}, the vanishing of $\lambda$ and $\lambda'$ was achieved through the fine-tuning of independent parameters, similar to the one necessary to achieve doublet-triplet splitting in the Higgs sector. In ref.~\cite{Tamvakis:1996np}, a mechanism similar to the missing-partner solution of the 2--3 splitting in SU(5)~\cite{Masiero:1982fe,Grinstein:1982um} was considered, at the price of introducing a number of relatively large representations. 
In this paper we will show that the superpotential in \eq{eq:Weffren} can be obtained without the need of fine-tuning in a relatively simple SO(10) model involving only fundamental, spinorial, and adjoint representations, thanks to the vacuum expectation value (vev) of an adjoint aligned along the $T_{3R}$ or $T_{B-L}$ direction. 

The paper is organized as follows. In \sect{frameworkAR} we illustrate the basic ingredients 
needed in order to generate baryonic RPV coupling in GUTs like SU(5) and SO(10). 
A class of explicit models is presented in \sect{expmodels}, 
while \sect{minimalSO10model} is devoted to a detailed analysis of a simple 
model where the vev of an adjoint is aligned along the $T_{3R}$ generator. 
In \sect{phenoremarks} we make some phenomenological remarks on the model 
and conclude in \sect{conclusions}. More technical details are collected in \apps{MS101616bar}{stopNLSP}.

\section{The framework}
\label{frameworkAR}

In this section, we define the rules of the game and systematically explore the options available in SU(5) and SO(10) to generate the superpotential in \eq{eq:Weffren}. The reader interested to specific models can jump to section \ref{minimalSO10model}. The main assumptions will be i) the use of representations that can arise in perturbative string theory \cite{Dienes:1996du}, ii) a renormalizable origin of the extra term in \eq{eq:Weffren}, and iii) the absence of fine-tuning. 

The basic idea which allows to generate the baryon number violating coupling 
$\lambda''$ at low energy, while conserving lepton number, is a split embedding of the MSSM 
fields into GUT representations. 
By a ``split embedding'' we mean that the MSSM chiral superfields usually embedded in a single 
GUT representation are embedded into distinct representations. 
Such a splitting can be realized as a consequence of the GUT-symmetry breaking in the presence of extra vectorlike representations.

Let us clarify this point with an example. 
In the ordinary embedding of SO(10), the MSSM fields belonging to one generation 
are contained in a $16$.
Let us imagine, however, to introduce a $10$ representation which mixes with $16$ 
through a term in the superpotential 
of the type
\begin{equation}
\label{splitembeddingexample1}
W = 16_H 16 \, 10 + \ldots \, , 
\end{equation}
where $16_H$ is an SO(10)-breaking field acquiring a GUT-scale vev, 
$V_{16} \equiv \vev{16_H}$, in the SU(5)-invariant direction. 
After substituting the vev into $16_H$ and decomposing \eq{splitembeddingexample1} under SU(5), we get 
\begin{equation}
\label{splitembeddingexample2}
W = V_{16} \, \overline{5}_{16} 5_{10} + \ldots \, , 
\end{equation}
namely $\overline{5}_{16}$ gets superheavy, while the SU(5) component 
$\overline{5}_{10}$, living in the 10, remains massless as long as there is no mass term for $10$ in the superpotential. 
Hence, the MSSM degrees of freedom (d.o.f.) are now 
split in an SU(5)-invariant way between the $16$, which ends up containing $q$, $u^c$, $e^c$, and 10, 
which ends up containing $d^c$, $l$.

Our mechanism to generate baryonic RPV relies on a similar split-embedding mechanism that, as we will see, will not be SU(5) preserving.

\subsection{SU(5)}

The case of SU(5) turns out not to offer any viable option. Still, it is useful to review it in order to illustrate the logic we will follow in this section, to find results that we will use in the next subsection, and to demonstrate that the fine-tuned method used in~\cite{Smirnov:1995ey} is the only way to obtain \eq{eq:Weffren} using only the representations $5$, $10$, $15$, $24$ (and conjugated, where relevant) available according to our assumptions.

To identify the renormalizable SU(5) origin of the operator $u^c_i d^c_j d^c_k$ ($i,j,k$ fixed and $j\neq k$), let us first observe that the light $u^c_i$ field must be contained in a 10 of SU(5), while $d^c_j$ and $d^c_k$ must be contained into two different $\overline{5}$, $\overline{5}'$ of SU(5), so that $u^c_i d^c_j d^c_k$ originates from the SU(5) operator $10\, \overline{5}\,\overline{5}'$. 

Let us denote by $L$, $L'$ the SU(5) partners of $d^c_j$, $d^c_k$ in $\overline{5}$, $\overline{5}'$, respectively, and by $E^c$, $Q$ the SU(5) partners of $u^c_i$ in 10. Then
\begin{equation}
\label{eq:SU5}
10\, \overline{5}\,\overline{5}' = u^c_i d^c_j d^c_k + E^c L L' +  Q d^c_j L' + Q L d^c_k . 
\end{equation}
For lepton number violating operators involving light fields not to be generated at the renormalizable level, at least two out of the four fields $L$, $L'$, $E^c$, $Q$ should not be light or partially light, in the sense that they should not contain the light fields $l_i$, $q_i$, $e^c_i$ even as a component. A splitting, analogous to the doublet-triplet splitting in the Higgs sector, must occur in either $\overline{5}$ or $\overline{5}'$ or 10. 

Let us first consider the case in which one of the two leptonic fields is heavy, say $L$ for definiteness, and denote by $\overline{5}_a$ the additional SU(5) representation containing the light lepton doublet $l_a$, $a=1,2,3$. Note that extra matter representations (four antifundamentals overall, $\overline{5}_1,\overline{5}_2,\overline{5}_3,\overline{5}$) are needed to realize a split embedding of the Standard Model (SM) fermions. To preserve the SM chirality content, one fundamental, $5$, must also be present, to compensate the  extra $\overline{5}$. A superheavy mass term is then allowed in the form
\begin{equation}
\label{eq:mass5}
5 (\mu_a + \alpha_a \vev{24_H}) \overline{5}_a ,
\end{equation}
where the $24_H$ is an SU(5) adjoint getting vev along the hypercharge generator, $\vev{24_H} = V\, Y$. Now, our definitions and assumptions require $d^c_j$ to have a component in $\overline{5}$ and the doublets $l_a$ to be light. For the light $d^c_j$ to have a component in $\overline{5}$, the mass term arising from~\eq{eq:mass5} must be nonzero for some $a = 1,2,3$,
\begin{equation}
\label{mass3}
\mu_a + \frac{\alpha_a}{3} V \neq 0 \, , 
\end{equation}
otherwise the $d^c_a$ would also be fully contained in the $\overline{5}_a$. As a consequence, at least one of the two vectors $(\mu_a)_{a=1,2,3}$ and $(\alpha_a)_{a=1,2,3}$ should be nonvanishing. On the other hand, in order for the doublets $l_a$ to be light, with no heavy component, the leptonic mass term arising from~\eq{eq:mass5} must vanish, 
\begin{equation}
\label{mass2}
\mu_a - \frac{\alpha_a}{2} V = 0 \, .
\end{equation}
The two above relations imply a fine-tuning in the necessary alignment of the two nonvanishing vectors $(\mu_a)_{a=1,2,3}$ and $(\alpha_a)_{a=1,2,3}$ and in the determination of the vev $V$. The argument easily generalizes to the case of more than two extra $5 \oplus \overline{5}$ or more than an adjoint getting a vev. 

The argument above also applies to the case in which neither $L$ nor $L'$ is fully heavy. In such a case, $Q$ and $E^c$ should both be, in order to prevent lepton number violating operators involving light fields to be generated. And again a splitting must be arranged between $u^c_i$ and its SU(5) partners, $Q$ and $E^c$, such that $u^c_i$ ends up having a vanishing mass. Since the only source of SU(5) breaking available, the vev of the SU(5) adjoints, never vanishes on the $L$, $L'$, $E^c$, $Q$ fields, a fine-tuned cancellation with another mass term must be invoked. In principle such a cancellation could be forced to arise dynamically, as in the sliding singlet solution of the 2-3 splitting problem~\cite{Barr:1997pt}, but this does not seem to be trivially possible in SU(5). 

The above discussion identifies two important ingredients  to obtain baryonic RPV in a natural way: i) a source of SU(5) breaking splitting the mass of some unified multiplets in such a way that a component remains massless, i.e.\ a source of SU(5) breaking projecting out some components of a unified multiplet; and ii) additional (vectorlike) matter, in order to be able to realize a split embedding of the SM fermions. SU(5) misses the first ingredient, which is, however, available in SO(10). 

\subsection{SO(10)}

In the case of SO(10), the available nontrivial representations are 10, 16, $\overline{16}$, 45, 54. The fields $u^c$ can be contained in the representations 16 and 45, while the fields $d^c$ can be contained in the representations 16 and 10. Therefore, the only SO(10)-invariant renormalizable origins of the operator $u^c_i d^c_j d^c_k$ are $16\, 16' 10$ (where 16 and $16'$ can coincide) and $45\, 10 10'$ (where $10$ and $10'$ must be different). 

In both cases, the embedding of $u^c_i$ proceeds through a 10 of SU(5), and the embedding of $d^c_j$ and $d^c_k$ proceeds through a $\overline{5}$ and $\overline{5}'$ of SU(5), respectively. The operator $u^c_i d^c_j d^c_k$ then again arises  from the SU(5) operator $10\, \overline{5}\,\overline{5}'$ appearing in the decomposition of both $16\, 16' 10$ and $45\, 10\, 10'$. We can then conclude that in both cases the decomposition of the SO(10) operator will contain the rhs of \eq{eq:SU5}, where we have denoted with $L$, $L'$, $E^c$, $Q$ the SU(5) partners of $d^c_j$, $d^c_k$, $u^c_i$ in $\overline{5}$, $\overline{5}'$, 10, as before. Again, at least two out of the fields $L$, $L'$, $E^c$, $Q$ must not contain a light component. 

Let us again first suppose that one of the two heavy fields is a lepton doublet, say $L$ for definiteness. Then the light (SM) leptons $l_a$, $a=1,2,3$, should be contained in three $\overline{5}_a$ independent of $\overline{5}$. We then have at least four antifundamentals of SU(5), which means that at least one fundamental of SU(5), $5$, must exist as well, with the mass mixing $5\,\overline{5}_a$ nonvanishing for the coloured components (otherwise, the light $d^c_a$ would be entirely contained in the $\overline{5}_a$, with no component in the $\overline{5}$) but vanishing for the lepton components (because the $l_a$ must be entirely contained in the $\overline{5}_a$, with no component in the $\overline{5}$). 

Unlike SU(5), SO(10) offers the possibility to achieve such a splitting without fine-tuning. As argued, a source of SU(5) breaking vanishing on the lepton components is needed. With the available field content, such a source can only be provided by the appropriately oriented vev of an adjoint. More precisely, there are two options, depending on the SO(10) operator from which the mass mixing $5\,\overline{5}_a$ arises (which for simplicity we assume to be  the same for the three families): 

\begin{itemize}
\item
If the operator originates from the SU(5) fundamental and antifundamental 
components of a $\overline{16}$ and three $16_a$, a mass term mixing the coloured components of $5$ and $\overline{5}_a$, but not the lepton ones, can be obtained through the SO(10) interaction
\begin{equation}
\label{T3Rmethod}
\alpha_a \overline{16}\, 45_H 16_a \, , 
\end{equation}
with the SO(10) adjoint $45_H$ getting a vev $\vev{45_H} = V_{45} T_{3R}$ along the 3R direction. Such a vev can be obtained without fine-tuning 
in a number of ways~\cite{Babu:1994dc,Dvali:1996wh}. 
\item
If the operator originates from the SU(5) fundamental and antifundamental 
of a $10$ and three $10_a$, a mass term mixing the coloured components of $5$ and $\overline{5}_a$, but not the lepton ones, can be obtained through the SO(10) interaction
\begin{equation}
\label{TBLmethod}
\alpha_a 10\, 45_H 10_a \, , 
\end{equation}
with the SO(10) adjoint $45_H$ getting a vev $\vev{45_H} = V_{45} T_{B-L}$ along the B-L direction. Such a vev can also be obtained without fine-tuning in a number 
of ways~\cite{Babu:1994dc,Dvali:1996wh}. 
\end{itemize}
In the next section, we will see that both the options can be implemented in the context of simple, minimal models.\footnote{In the complete models, the $\overline{5},\overline{5}_a$ defined in the SU(5) subsection end up being superpositions of the antifundamentals in $16_a,16$ or $10_a,10$.}

So far we have assumed that at least one of the two heavy fields among $L$, $L'$, $E^c$, $Q$ is a lepton doublet. Let us now assume that this is not the case. Then, both $E^c$ and $Q$ should be fully heavy. And the light (SM)  $e^c_a$, $q_a$, $a=1,2,3$ should be contained in three $10_a$ of SU(5), independent of the $10$ containing $u^c_i$. We then have at least four 10 of SU(5), which means that at least one $\overline{10}$ must exist, with the mass mixing $\overline{10}\, 10_a$ vanishing  
for the lepton singlet and quark doublet components but nonvanishing on the quark singlet components. Unfortunately, not even SO(10) allows us to achieve such a splitting without fine-tuning, independently of whether the $10_a$ of SU(5) are embedded in spinorial or adjoint representations of SO(10). Therefore, the cases considered above are the only relevant ones. 

\section{Explicit models}
\label{expmodels}

In this section we discuss simple, minimal realizations of the two basic mechanisms outlined in the previous section to obtain~\eq{eq:Weffren}. In both cases, the RPV operator will arise from the decomposition of an SO(10) operator in the form $16\, 16' 10$ (where $16$ and $16'$ may or may not coincide). Models in which RPV arises from an operator in the form $45\, 10\, 10'$ are also possible, but since they involve a larger number of fields, we will not present them here. 

The vev of a $45_H$ along the $T_{3R}$ or $T_{B-L}$ direction can be obtained as in~\cite{Babu:1994dc,Dvali:1996wh} through an SO(10) breaking sector that also generates a vev for a $16_H \oplus \overline{16}_H$ along the SM-singlet direction, as necessary to fully break SO(10) to the SM. A renormalizable superpotential $W_H$, 
also involving a $54_H$ and an SO(10) singlet, is sufficient to achieve such vevs. The SO(10) breaking fields above will always appear together with two ``matter fields'' in the rest of the superpotential, which guarantees that the supersymmetric minimum provided by $W_H$ is not affected by the rest of the superpotential. 

\subsection{Adjoint vev along the $T_{3R}$ direction}
\label{SO10T3R}

In this case, the operator relevant for the necessary splitting of leptons and baryons is $\alpha_a \overline{16}\, 45_H 16_a$, with $45_H$ assumed to get a vev $\vev{45_H} = V_{45} T_{3R}$ in the $T_{3R}$ direction. On top of the three $16_a$ needed to reproduce the SM chiral field content, the ``matter'' content  necessarily involves a 
$16 \oplus \overline{16}$ and a 10 (the latter in order to be able to write a RPV source in the form $16\,16\,10$). As mentioned, the SO(10)-breaking sector must involve a $16_H \oplus \overline{16}_H$ getting vev along the SM-singlet components. The case in which the role of $16_H \oplus \overline{16}_H$ is played by $16 \oplus \overline{16}$ can be in principle considered, but here we will assume for simplicity that this is not the case. 
The minimal matter content relevant to our goal, which for the time being is to generate the RPV source, is then
\begin{equation}
\label{eq:matter1}
16_a,
16, 
\overline{16}, 
10 \quad
45_H, 
16_H,
\overline{16}_H.
\end{equation}
Accounting for the SM Higgs and Yukawas needs an additional $10_H$, as we will discuss below.
The three possible sources of the RPV operator $u^c_i d^c_j d^c_k$ are $16\, 16\, 10$, $16_a 16\, 10$, $16_a 16_b 10$. The last one is not ideal, as it generically also generates lepton number violating operators, unless a specific flavour structure is specified. On the other hand, it is relatively easy to use $16\, 16\, 10$ or $16_a 16\, 10$. In both cases the superpotential leading, at low energy, to \eq{eq:Weffren}, is essentially unique. 

If the RPV operator originates from $16\, 16\, 10$, we are led to a superpotential in the form
\begin{multline}
\label{eq:W1}
W_1 = \lambda 16\, 16\, 10 + \alpha_a \overline{16}\, 45_H 16_a \\ + \beta_a 16_H 16_a 10 + M_{16}\overline{16} 16 \, .
\end{multline}
The RPV operator arises from $16\,16\,10$ because of the mixing between $16_a$, 16, 10 induced by the terms  $\alpha_a \overline{16}\, 45_H 16_a$ and  $\beta_a 16_H 16_a 10$ after SO(10) breaking. The first term only affects the singlet fields $u^c$, $d^c$, $e^c$, while the second term only affects the $d^c$, $l$ fields. The light quark doublets $q_a$ are not mixed by either operators, and therefore lie in the $16_a$. One lepton doublet acquires a component in the 10 because of the $\beta_a \vev{16_H} 16_a 10$ mixing. One lepton singlet and one up quark singlet acquire a component in the 16 because of the $\alpha_a \overline{16}\, \vev{45_H} 16_a$ mixing. The down quark singlets spread in the $16_a$, $16$, and $10$ as they are affected by both mixing terms. See also \Table{tab:fieldcontentW1} for a summary of the MSSM embedding 
resulting from $W_1$ in \eq{eq:W1}.
As a consequence, the operators $q_i d^c_j l_k$ and $e^c_i l_j l_k$ are not generated by $16\, 16\, 10$, while $u^c_i d^c_j d^c_k$ are. A more detailed discussion can be found in Appendix~\ref{MS101616bar}.

\begin{table}[h]
\begin{tabular}{|c|c|c|c|c|c|}
\hline
 & $q$ & $u^c$ & $d^c$ & $l$ & $e^c$ \\
\hline
$16_a$ &  $\checkmark$ &  $\checkmark$ &  $\checkmark$ &  $\checkmark$ &  $\checkmark$  \\
\hline
$16$ &  &  $\checkmark$ &  $\checkmark$ &  &  $\checkmark$  \\ 
\hline
$10$ &  &  & $\checkmark$ & $\checkmark$ &   \\
\hline
\end{tabular}
\caption{The SO(10) matter fields and the light MSSM components they contain 
for the case of the superpotential $W_1$ in \eq{eq:W1}.}
\label{tab:fieldcontentW1}
\end{table}

Notice that the two vectors $\alpha_a$ and $\beta_a$ need to be linearly independent in order to obtain $\lambda''_{ijk}\neq 0$. This can be seen as follows. If $\alpha_a$ and $\beta_a$ were parallel, it would be possible to choose a basis for the $16_a$ such that $\alpha_{1,2} = \beta_{1,2} = 0$. In such a basis, the first two families of the light fermions are contained in $16_{1,2}$ and only the third family mixes with $16$ and $10$. There is therefore only a single light eigenstate $d^c_l$ with components in both 16 and 10. The coupling $\lambda''_{ijk}$ then vanishes because the antisymmetry in $j,k$ requires two different light eigenstates to have components in 16 and 10. Another way of rephrasing this result is that $\lambda''_{ijk}$ vanishes in the U(2)-symmetric limit, where U(2) acts on $16_{1,2}$~\cite{Pomarol:1995xc,Barbieri:1995uv,Barbieri:1996ww,Barbieri:1997tu}. If the size of U(2) breaking is set by the light Yukawa couplings of the SM, baryonic RPV will necessarily end up being correspondingly suppressed. 

There is no room for a light Higgs field with the spectrum in \eq{eq:matter1} and the superpotential in \eq{eq:W1}. 
An additional $10_H$ must therefore be added in order to accommodate it. The MSSM Yukawas are then generated by terms in the form $y 16 16 10_H$ or $y_a 16_a 16 10_H$ or $y_{ab} 16_a 16_b 10_H$. Doublet-triplet splitting should be accounted for separately, 
but all the ingredients for the Dimopoulos-Wilczek mechanism are 
available \cite{Dimopoulos:1981xm,Babu:1993we,Babu:1994dq,Babu:1994dc,Berezhiani:1996bv,Barr:1997hq,Chacko:1998jz,Maekawa:2001uk,Babu:2010ej}. 

In \eq{eq:W1} we have included only interactions coupling $16_H$, $\overline{16}_H$, $45_H$ to two matter fields, as anticipated. A mass term in the form $\overline{16} 16_a$ can be eliminated by an SU(4) rotation of the four spinorials $16$, $16_a$, $a=1,2,3$. Possible $\lambda_a 16_a 16\, 10$ and $\lambda_{ab} 16_a 16_b 10$ terms are not allowed as  they would give rise to $q\, d^c l$ operators. On the other hand, terms such as $\overline{16}_H\overline{16} \, 10$, $\overline{16}\, 45_H 16$, $M_{10}10^2$ would not modify our conclusions. 

The second case we consider is associated to the following superpotential:
\begin{multline}
\label{eq:W2}
W_2 = \lambda_a 16_a 16\, 10 + \alpha_a \overline{16}\, 45_H 16_a \\ +\beta 16_H 16 10 + \overline{\beta} \,\overline{16}_H \overline{16} 10 + M_{16}\overline{16} 16 \, .
\end{multline}
The RPV operator arises from $16_a16\,10$ because of the mixing between $16_a, 16, 10$ induced by the terms  $\alpha_a \overline{16}\, 45_H 16_a$ and  $\beta 16_H 16 10$ after SO(10) breaking. The light lepton and quark doublets are fully contained in the $16_a$, so that no lepton number violating operators can 
be generated (see also \Table{tab:fieldcontentW2}).
The two vectors $\alpha_a$ and $\lambda_a$ need to be linearly independent in order to obtain $\lambda''_{ijk}\neq 0$. 

\begin{table}[h]
\begin{tabular}{|c|c|c|c|c|c|}
\hline
 & $q$ & $u^c$ & $d^c$ & $l$ & $e^c$ \\
\hline
$16_a$ &  $\checkmark$ &  $\checkmark$ &  $\checkmark$ &  $\checkmark$ &  $\checkmark$  \\
\hline
$16$ &  &  $\checkmark$ &  $\checkmark$ &  &  $\checkmark$  \\ 
\hline
$10$ &  &  & $\checkmark$ & &   \\
\hline
\end{tabular}
\caption{Same as in \Table{tab:fieldcontentW1} for the superpotential $W_2$ in \eq{eq:W2}.}
\label{tab:fieldcontentW2}
\end{table}

The light Higgs could be in principle accommodated in the 10, $16_3$ and $\overline{16}$ (in the basis in which $\alpha_{1,2}=0$) and doublet-triplet splitting achieved for free if $\overline{\beta} = 0$. In such a case, however, the light down singlets would be contained in $16_{1,2}$ and $10$, and no down quark Yukawa would be generated. Therefore, we need to assume $\overline{\beta} \neq 0$ (or, equivalently, a nonvanishing mass term $M_{10}10^2$) and to add an additional $10_H$ to accommodate the light Higgs fields. The MSSM Yukawas are then generated by terms in the form $y 16 16 10_H$ or $y_a 16_a 16 10_H$ or $y_{ab} 16_a 16_b 10_H$. 

A mass term in the form $\overline{16} 16_a$ in \eq{eq:W2} can be eliminated by a SU(4) rotation of the four spinorials $16$, $16_a$, $a=1,2,3$. Possible $\beta_a 16_H 16_a 10$ and $\lambda_{ab} 16_a 16_b 10$ terms are not allowed as  they would give rise to $q\, d^c l$ operators. The presence of the terms $\lambda 16\,16\,10$, $\overline{\lambda} \overline{16}\, \overline{16}\, 10$, $\alpha \overline{16} 45_H 16$ would not affect the conclusions above. 

\subsection{Adjoint vev along the $T_{B-L}$ direction}

In this case, the operator relevant for the necessary splitting of leptons and baryons in the unified multiplets is $\alpha_a 10\, 45_H 10_a$, with $45_H$ assumed to get a vev $\vev{45_H} = V_{45} T_{B-L}$ in the $T_{B-L}$ direction. On top of the three $16_a$ needed to reproduce the SM chiral field content, the ``matter'' content   involves a 10 and three $10_a$, $a=1,2,3$. The minimal matter content relevant to our goal is then
\begin{equation}
\label{eq:matter2}
16_a,
10_a, 10 \quad
45_H, 
16_H,
\overline{16}_H.
\end{equation}
The  possible sources of the RPV operator $u^c_i d^c_j d^c_k$ are $16_a 16_b 10$, $16_a 16_b 10_c$. The latter  generically also generates lepton number violating operators, unless a specific flavour structure is specified. Let us then consider the following superpotential involving the former:
\begin{multline}
\label{eq:W3}
W_3 = \lambda_{ab} 16_a 16_b 10 +\alpha_a 10\, 45_H 10_a \\+\alpha_{ab} 10_a 45_H 10_b + h_{ab} 16_H 16_a 10_b. 
\end{multline}
The light fields $q_a$, $u^c_a$, $e^c_a$ are only contained in the $16_a$. The operator $h_{ab} 16_H 16_a 10_b$ forces the light lepton doublets $l_a$ to lie in the $10_a$ only, whereas the light $d^c_i$ are both in the $10_a$, the $16_a$, and the $10$ because of the mixing induced by $\alpha_a 10\, 45_H 10_a$ and $\alpha_{ab} 10_a 45_H 10_b$ (note that the second one is necessary; otherwise, only a single light component would appear in both $16_a$ and $10$, and $\lambda''_{ijk}$ would vanish because of the antisymmetry). 
See also \Table{tab:fieldcontentW3} for a summary of the MSSM embedding 
resulting from $W_3$ in \eq{eq:W3}.
Only the lepton number conserving RPV operator is thus generated by the $16_a 16_b 10$ term. 

\begin{table}[h]
\begin{tabular}{|c|c|c|c|c|c|}
\hline
 & $q$ & $u^c$ & $d^c$ & $l$ & $e^c$ \\
\hline
$16_a$ &  $\checkmark$ &  $\checkmark$ &  $\checkmark$ &  &  $\checkmark$  \\
\hline
$10_a$ &  &  &  $\checkmark$ & $\checkmark$ &   \\ 
\hline
$10$ &  &  & $\checkmark$ & &   \\
\hline
\end{tabular}
\caption{Same as in \Table{tab:fieldcontentW1} for the superpotential $W_3$ in \eq{eq:W3}.}
\label{tab:fieldcontentW3}
\end{table}

The embedding of the $l_a$ and part of the $d^c_a$ in the $10_a$, forced by the operator $h_{ab} 16_H 16_a 10_b$ allows to obtain positive, universal sfermion masses at the tree level, if supersymmetry is broken by the vev of a 16~\cite{Nardecchia:2009ew,Nardecchia:2009nh,Monaco:2013poa,Caracciolo:2012de,Caracciolo:2012jq}. In this context, the presence of three $16_a \oplus 10_a$ can be associated to a further stage of unification in $E_6$~\cite{Monaco:2011fe}. The embedding through $16_a \oplus 10_a$ also allows to obtain a predictive framework for leptogenesis~\cite{Frigerio:2008ai,Calibbi:2009wk}. 

The doublet-triplet splitting in the Higgs sector could be in principle obtained for free. 
Indeed, the doublets in the 10 are also light and could play the role of the Higgs doublets. 
The up Yukawa interactions would in this case be provided by the very same operator generating $\lambda''_{ijk}$. 
However, no lepton Yukawa would be generated. Therefore, we need to add again an additional $10_H$ to accommodate the light Higgs fields and make the doublets in the 10 heavy by adding a $M_{10} 10^2$ mass term. 

Adding the term $\lambda_{abc} 16_a 16_b 10_c$ or mass terms in the form $M_{ab} 10_a 10_b$ or $M_a 10_a 10$ would introduce lepton number violation. All other terms involving two matter fields are allowed. 

\subsection{On the structure of the superpotential}
\label{naturalnessW}

A comment on the flavour structure of the superpotential in \eq{eq:W1} is in order. We achieved our goal of generating an isolated baryonic RPV operator without invoking a special structure with respect to the flavour index $a = 1,2,3$. On the other hand, we implicitly distinguished the three $16_a$ from the $16_H$ and $16$. For example, we assumed $16_H 16_a 10$ to be present in the superpotential in \eq{eq:W1}, while $16_a 16_b 10$ is not. The question then arises whether it is possible to find a symmetry forcing the superpotential to have the desired form. The answer to this question depends on the form of $W_H$, which constrains the quantum numbers of $16_H$, $\overline{16}_H$, $45_H$. Let us consider, for example, the case in which $W_H$ contains the terms $M_{45} 45^2_H$ and $X(16_H\overline{16}_H - V^2_{16})$, where $X$ is an SO(10) singlet, as e.g.~in~\cite{Babu:1994dc,Nardecchia:2009nh}. 
In such a case, it turns out that it is not possible to find a symmetry that allows all the terms we need and forbids the ones that should not appear. In particular, it is not possible to find any symmetry that distinguishes  the fields $16_H$ and $16_a$.\footnote{On the other hand, it is possible to find a $Z_2$ symmetry which discriminates $16$ from $16_H$ and $16_a$ (and $\overline{16}$ from $\overline{16}_H$ as well). An explicit example is $Z_2(45_H,10,16_a,16,16_H,\overline{16},\overline{16}_H,X) = (-,+,+,-,+,-,+,+)$.} Nonetheless, the structure of the superpotential we need can be justified at a more fundamental level, once the origin of the flavour structure of the superpotential (and of the SM fermions) is addressed. For instance, one could envisage the presence of an ${\rm SU(3)}_H$ horizontal symmetry under which the $16_a$ transforms as the fundamental of ${\rm SU(3)}_H$, while the $16_H$ transforms trivially. The flavor symmetry is then formally restored in the superpotential considering the various couplings as spurions. We will illustrate this point in more detail in \sect{flavormodel}.

\section{Analysis of a simple model}
\label{minimalSO10model}

In this section we study in greater detail the first model of \sect{SO10T3R}. 
To this end we consider the superpotential in \eq{eq:W1} augmented with a mass term for the 
$10$. As anticipated in \sect{SO10T3R}, an additional $10_H$ 
must be added as well in order to accommodate the light Higgs doublets and the Yukawa sector.
The total superpotential is therefore 
\begin{equation}
\label{totalW}
W = W_\text{RPV} + W_Y \, , 
\end{equation}
where 
\begin{multline}
\label{WRPV1616bar10}
W_\text{RPV} = \lambda 16\, 16\, 10 + \alpha_a \overline{16}\, 45_H 16_a \\ 
+ \beta_a 16_H 16_a 10 
+ M_{16}\overline{16} 16 + \frac{M_{10}}{2} 10 \, 10 
\end{multline}
and 
\begin{equation}
\label{Wyuk10}
W_Y = y_{ab} 16_a \, 16_b \, 10_H + y_a 16_a 16 \, 10_H + y \, 16 \, 16 \, 10_H \, .
\end{equation}   
On top of that, we will assume an extra sector responsible for the 
alignment of the adjoint vev along the 3R generator and for the DT splitting. 
Not all terms allowed by the SO(10) symmetry are included in the superpotential in \eq{totalW}. 
Possible mechanisms to forbid such terms will be discussed in the next subsection. For simplicity, all the parameters are taken to be real.

In what follows, we will consider two different limits in the parameter space of this model. 
The first one is useful because it is particularly simple, and as such it allows to illustrate some features of the general results, 
though it does not lead to a realistic pattern of fermion masses and mixings as we will show below. 
The second limit, to be considered in \sect{flavormodel}, is interesting because, within motivated assumptions, it allows to understand features of the SM third-family fermion spectrum. 

Let us begin then by considering the limit in which the extra vectorlike states $10 \oplus 16 \oplus \overline{16}$ are much heavier than the the GUT vevs,
\begin{equation}
\label{eq:scales2}
M_{10},M_{16} \gg V_{16}, V_{45}.
\end{equation}
In such a case, the light MSSM superfields are mostly contained (up to $V / M$ corrections) in the $16_a$, and one can integrate out the heavy fields $10$, $16$, and $\overline{16}$ at the SO(10) level,
thus obtaining at the leading order in $1 / M$
\begin{align}
\label{10int}
10 & \approx - \frac{1}{M_{10}} \left( \beta_a 16_H 16_a \right)_{10} \, , \\
\label{16int}
16 & \approx - \frac{1}{M_{16}} \left( \alpha_a 45_H 16_a \right)_{16} \, ,
\end{align}
where the subscripts denote the proper SO(10) contractions and the $\overline{16}$ should be 
set to zero at this order. 
Substituting the full solutions for $10,16,\overline{16}$ into \eq{WRPV1616bar10} and expanding at the third order in $1/M$, we get
\begin{multline}
\label{WeffSO10A}
W^{\text{eff}}_\text{RPV} \approx - \frac{1}{2 M_{10}} \left( \beta_a 16_H 16_a \right)_{10}^2 \\
- \frac{1}{M_{16}^2 M_{10}} \lambda  \left( \alpha_a 45_H 16_a \right)^2_{16} \left( \beta_c 16_H 16_c \right)_{10} \, .  
\end{multline}
While the first term in \eq{WeffSO10A} is irrelevant for our purpose, 
the second one leads, upon GUT-symmetry breaking, to the $\Delta B = 1$ RPV operator $\lambda''_{abc} u^c_a d^c_b d^c_c$, with 
\begin{equation}
\label{lambdaseffSO10A}
\lambda''_{abc} = \frac{V_{45}^2 \, V_{16}}{M_{16}^2 \, M_{10}} \, \lambda \, \alpha_a \alpha_{[b} \beta_{c]} \, .
\end{equation}
In the expression above, the square brackets denote antisymmetrization.

\begin{figure}[ht]
\includegraphics[width=7.5cm]{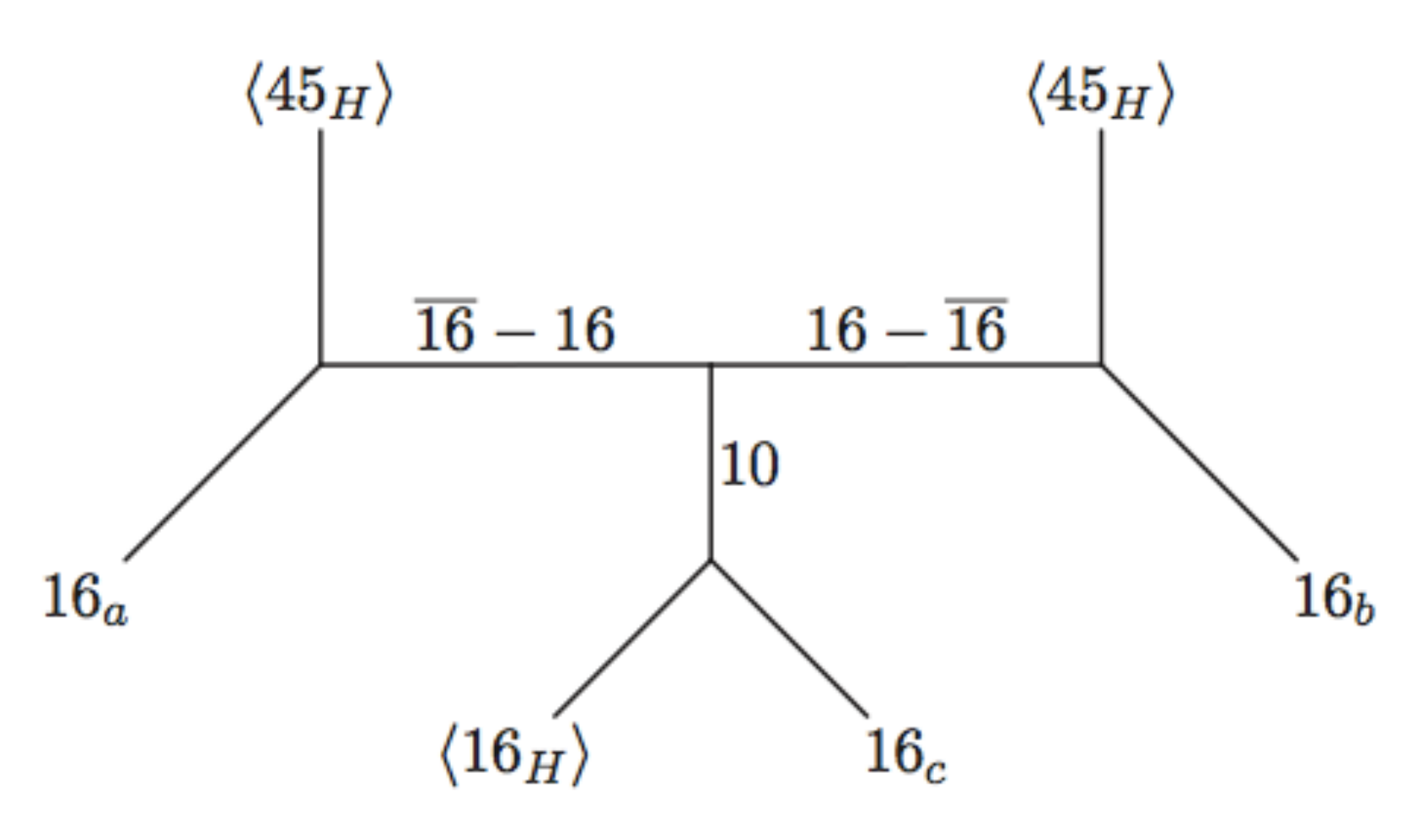}
\caption{\label{SO10A1} 
SO(10) superdiagram leading to the $\Delta B = 1$ RPV operator in the effective MSSM theory. 
The vertices and propagators are specified by the superpotential in \eq{WRPV1616bar10}. 
} 
\end{figure}

The result in \eq{lambdaseffSO10A} can be derived in a number of different ways. 
For instance, one can directly inspect the mass matrices of the relevant fields upon GUT-symmetry breaking 
(cf.~\eq{lambdasdeclim} in \app{MS101616bar}) 
or, from a diagrammatic point of view, compute the tree-level graph in \fig{SO10A1}.

Note that the light fields $u^c_a$, $d^c_a$ in \eq{lambdaseffSO10A} do not necessarily correspond to fermion mass eigenstates. The latter are in fact determined by the diagonalization of the SM Yukawa couplings, which have not been specified so far. In the fermion mass eigenstate basis, in which the low-energy bounds on $\lambda''_{ijk}$ are extracted, \eq{lambdaseffSO10A} becomes 
\begin{equation}
\label{lambdasFS}
\lambda''_{ijk} \propto (V_{u^c})_i^a  (V_{d^c})_{[j}^b (V_{d^c})_{k]}^c 
\alpha_a \alpha_b \beta_c \, ,
\end{equation}
where $V_{u^c}$ and $V_{d^c}$ are the unitary transformations used to diagonalize the up and down Yukawa couplings (on the quark singlet side), determined by the SO(10) Yukawa sector. 

This leads us to the discussion of the Yukawa sector. 
Simple expressions for the SM Yukawa matrices can be obtained at the leading order in the limit $M\gg V$. 
Substituting \eq{16int} into the superpotential $W_Y$ in \eq{Wyuk10}, we get:
\begin{equation}
\label{WeffYuk}
\!\!\! W^{\rm{eff}}_Y = 
y_{ab} 16_a \, 16_b \, 10_H 
- \frac{y_a}{M_{16}} 16_a \left( \alpha_b 45_H 16_b \right)_{16} 10_H \, ,
\end{equation} 
where the last term in \eq{Wyuk10} has been neglected since it does not to contribute to SM fermion masses. 
 
Denoting the up-quark, down-quark, charged-lepton, and Dirac-neutrino mass matrices by 
$M_u$, $M_d$, $M_e$, and $M_D$, respectively, \eq{WeffYuk} leads to 
\begin{align}
\label{Mudeclim}
(M_u)_{ab} &= (2y_{ab} + \theta \, y_a \hat{\alpha}_b) v_u \, , \\
\label{Mddeclim}
(M_d)_{ab} &= (2y_{ab} - \theta \, y_a \hat{\alpha}_b) v_d \, , \\ 
\label{Medeclim}
(M_e)_{ab} &= (2y_{ab} - \theta \, y_a \hat{\alpha}_b) v_d \, , \\
\label{MDdeclim}
(M_D)_{ab} &= (2y_{ab} + \theta \, y_a \hat{\alpha}_b) v_u \, , 
\end{align}
where $y_{ab}$ is symmetric, $\theta \equiv  \alpha V_{45} / M_{16}$, $\alpha \equiv \sqrt{\sum_a \alpha_a^2}$, 
and $v_{u,d}$ are the electroweak vevs. 
Note that the relation $M_u = M_D$ implies that the neutrino sector must be extended with a 
Majorana mass term for $\nu^c \nu^c$. This can be achieved, for instance, by means of the effective 
operator $16_a 16_b \overline{16}_H \overline{16}_H / \Lambda$.\footnote{In this context it is worth it to recall 
that, due to the selection rules imposed by kinematics and Lorentz invariance, 
the simultaneous presence of $\Delta B = 1$ and $\Delta L = 2$ interactions do not 
endanger matter stability.}
The superpotential in \eq{Wyuk10}, complemented with the effective neutrino-mass operator, 
can reproduce the observed patter of fermion masses and 
mixings, but the larger hierarchy of masses in the up sector and the 
deviations from SU(5) relations for the light down quark and charged lepton require a certain amount of fine-tuning. Moreover, the above equations do not address the origin of the fermion mass hierarchy. Both such issues can be addressed in the context of flavour models, as shown by the simple example in the next subsection. 

\subsection{Addressing flavour}
\label{flavormodel}

So far, we did not make any assumption on the flavour structure of the couplings in \eq{WRPV1616bar10}. On the other hand, the latter is relevant for 
a number of reasons: to account at the same time for the pattern of SM fermion masses and mixings, 
to distinguish different representations with the same gauge quantum number (e.g.~$16_H$ and $16_a$), thus making the superpotential in \eq{WRPV1616bar10} technically natural, and to relate the size of the RPV couplings to the pattern of fermion masses and mixings. In this section we analyze the consequences of having a controlled flavour structure by means of a simple flavour model.  

Let us assume that the theory specified by \eq{WRPV1616bar10} and \eq{Wyuk10} is invariant under the horizontal symmetry group ${\rm SU(3)}_H$,\footnote{The horizontal ${\rm SU(3)}_H$ symmetry in the context of GUTs was originally discussed 
in Refs.~\cite{Berezhiani:1983rk,Berezhiani:1983hm,Berezhiani:1985in}.} 
with the $16_a$ transforming as a triplet and all the other fields transforming trivially. 
Let us also assume then that the horizontal symmetry is broken by the vev of two linearly independent spurion fields $A$ and $B$, which transform as antitriplets of ${\rm SU(3)}_H$ and whose absolute values are hierarchical, $\left| A \right| \gg \left| B \right|$. We neglect the masses and mixings related to the first families, which are zero in the absence of a third source of ${\rm SU(3)}_H$ breaking.

With $A$ and $B$ being the only sources of flavour symmetry breaking, 
we can write the parameters $\alpha_a, \beta_a, y_a, y_{ab}$ in terms of the spurions $A_a$ and $B_a$, in such a way that the superpotential in \eq{WRPV1616bar10} and \eq{Wyuk10} is formally invariant under the horizontal ${\rm SU(3)}_H$,
\begin{align}
\alpha_a &= r_{\alpha} A_a + s_{\alpha} B_a \, , \\
\beta_a &= r_{\beta} A_a + s_{\beta} B_a \, , \\
y_a &= r_{z} A_a + s_{z} B_a \, , \\
y_{ab} &=  r_{y} A_a A_b + s_{y} B_a B_b + t_y \left( A_a B_b + B_a A_b \right)  \, ,
\end{align}
where the coefficients $r_{\#}$, $s_{\#}$, and $t_y$ are $\mathcal{O}(1)$ numbers, but they could be assumed to be small or vanishing without fine-tuning. For the same reason, unwanted interactions such as $\lambda_a 16_a 16\, 10$ can be assumed to be absent from \eq{WRPV1616bar10} without fine-tuning. 

In what follows it turns out to be  useful to trade the vectors $A_a$ and $B_a$ for $\alpha_a$ and $\beta_a$ and, by means of an 
${\rm SU(3)}_H$ rotation, to go in the basis $(\alpha_a) = \alpha (0,0,1)$ and $(\beta_a) = \beta (0,\epsilon, 1)$,
where $\alpha$ and $\beta$ are $\mathcal{O}(1)$ numbers and $\epsilon \ll 1$, as a consequence of $|B|\ll|A|$. 
In the latter basis, the remaining parameters of the superpotential transforming nontrivially under the flavour group 
are
\begin{align}
\label{y33eps}
y_{33} \sim y_3 & = \mathcal{O} (1) \, , \\[0.8mm]
\label{y23eps}
y_{23} = y_{32} \sim y_2 & = \mathcal{O} (\epsilon) \, , \\
\label{y22eps}
y_{22} & = \mathcal{O} (\epsilon^2) \, .
\end{align}
For simplicity we shall factor out the appropriate $\epsilon$ dependence from the parameters in \eqs{y23eps}{y22eps}, i.e.~$y_{23} \to y_{23} \epsilon$, $y_{2} \to y_{2} \epsilon$ and $y_{22} \to y_{22} \epsilon^2$, so that all the parameters of the superpotential except $\epsilon$ are $\mathcal{O}(1)$ numbers. 

At this point one can inspect the mass matrices after SO(10)-symmetry breaking from \eq{WRPV1616bar10} 
and find the light MSSM content of $16_a$, $16$, $10$ (cf.~\eqs{dclikeproj}{qlikeproj} in \app{detflavmod}). 
The Yukawa matrices (in the $2\times 2$ approximation) can then be read directly from \eq{Wyuk10}. 
We report them for completeness in \eqs{Yuflavormod}{Yeflavormod} of \app{detflavmod}.
At the leading order in $\epsilon$, they yield the relations for the physical observables 
\begin{align}
\label{mtfm}
m_t &= (2 c_{\theta} y_{33} + s_\theta y_3 )v_u \, , \\[1mm]
\label{mcfm}
m_c &= \epsilon^2 \left(2 y_{22} - 2y_{23}\frac{2c_\theta y_{23}+ s_\theta y_2}
{2c_\theta y_{33}+ s_\theta y_3} \right) v_u \, , \\[1mm]
\label{mbfm}
m_b &= N \left(2 c_\theta y_{33}- s_{\theta} y_3 \right) v_d \, , \\[1mm]
\label{msfm}
m_s &=  \epsilon^2 \left( 2y_{22}  - 2 y_{23} \frac{2 c_\theta y_{23} -  s_{\theta} y_2}{2 c_\theta y_{33} -  s_{\theta} y_3}  \right) v_d\, , \\[1mm]
\label{mtaufm}
m_{\tau} &= c_\phi \left( 2 c_\theta y_{33} -  s_{\theta} y_3 \right) v_d  \, , \\[1mm]
\label{mmufm}
m_{\mu} &=   \epsilon^2 \left( 2y_{22} - 2 y_{23} \frac{2 c_\theta y_{23} - s_\theta y_2}{2 c_\theta y_{33} - s_\theta y_3} \right) v_d \, ,  \\[1mm]
\label{VCKMfm}
\left| V_{ts} \right| &= \left| V_{cb} \right| = \epsilon \left| 
\frac{2 c_\theta y_{23}+ s_\theta y_2}{2 c_\theta y_{33}+ s_\theta y_3} 
-\frac{2 c_\theta y_{23} - s_{\theta} y_2}{2 c_\theta y_{33} - s_{\theta} y_3} \right|  \, ,
\end{align}
where we defined the quantities:
\begin{equation}
\label{deftttpN3}
t_\theta \equiv \frac{V_{45} \alpha}{M_{16}} \, , \ \
t_\phi \equiv \frac{V_{16} \beta }{M_{10}} \, , \ \
N \equiv \left( \frac{1 + t_{\theta}^2}{1 + t_{\theta}^2 + t_{\phi}^2} \right)^{\tfrac{1}{2}} \, , 
\end{equation}
with $t$, $s$, and $c$ denoting the $\tan$, $\sin$ and $\cos$ functions, respectively. 

The expression above shows that the larger hierarchy in the up sector, $(m_c/m_t)_\text{GUT} \ll (m_s/m_b)_\text{GUT}$ at the GUT scale, can be due to $N\ll 1$ (so that a cancellation between the two terms in $2 c_\theta y_{33} - s_\theta y_3$ does not need to be invoked). Moreover, $(m_b)_\text{GUT} \approx (m_\tau)_\text{GUT}$ follows from $N\approx c_\phi$. The two conditions are both satisfied if $t^2_\theta \ll 1\ll t^2_\phi$, i.e.\ if
\begin{equation}
\label{eq:scales}
M_{10} < V_{16}, \quad V_{45} < M_{16},
\end{equation}
which can be interpreted as a sign of a two-step breaking $\text{SO(10)}\to \text{SU(5)}$ at the scale $V_{16} \sim M_{16}$ followed by $\text{SU(5)}\to G_\text{SM}$ at the lower scale $V_{45}\sim M_{10}$. 

On the other hand, the expressions in \eqs{mtfm}{mmufm} show that, independent of the limit chosen, $m_\mu \approx m_s$ at the GUT scale, which is not phenomenologically viable. This conclusion can be evaded if the subleading spurion $B$ is not SU(5) invariant (which may be associated to its being subleading). Let us then concentrate on the third-family relations. In the limit in \eq{eq:scales}, the expressions for the 
third-family fermion masses become
\begin{align}
\label{mtfm2}
m_t &\approx 2 y_{33} v_u \, , \\[2mm]
\label{mbfm2}
m_b &\approx 2y_{33}\left(\frac{M_{10}}{\beta V_{16}}\right) v_d \, , \\
\label{msfm2}
m_{\tau} &\approx 2y_{33}\left(\frac{M_{10}}{\beta V_{16}}\right) v_d  \, .
\end{align}

Let us now consider the size and the structure of the RPV couplings. The latter are obtained by projecting the $16\, 16\, 10$ operator in \eq{WRPV1616bar10} onto the light components (cf.~\eq{lambdasfmapp} in \app{detflavmod}) and by taking into account the subsequent 
electroweak rotation matrices $V_{u^c}$ and $V_{d^c}$ (cf.~\eqs{Vucflavormod}{Vdcflavormod} in \app{detflavmod}). This yields: 
\begin{align}
\label{lambdatbsfm}
\lambda''_{tbs} &= 2 \, \lambda \, \epsilon \, \frac{s_{\theta} t_{\theta} t_{\phi}}{(1 +t_{\theta}^2 + t^2_{\phi})^{1/2} } \, , \\
\label{lambdacbsfm}
\lambda''_{cbs} &= - \epsilon \, \frac{2 y_{23}}{2 c_\theta y_{33} + s_\theta y_3}  \lambda''_{tbs} \, .
\end{align}
The RPV couplings involving the first family vanish because, having introduced only two spurions, we have neglected the structure associated to the first-family masses. Note also that the RPV coupling is proportional to the small misalignment between the 3-vectors $\alpha_a$ and $\beta_a$, i.e. \eqs{lambdatbsfm}{lambdacbsfm} vanish in the $\epsilon \rightarrow 0$ limit. Eventually, we obtain a hierarchical structure for the RPV couplings.
In the limit in  \eq{eq:scales},  the expressions above simplify to
\begin{align}
\label{lambdatbsfm2}
\lambda''_{tbs} &= 2 \, \lambda \, \epsilon \, t_\theta^2 \, , \\[1mm]
\label{lambdacbsfm2}
\lambda''_{cbs} &= - \epsilon \, \frac{y_{23}}{y_{33}}  \lambda''_{tbs} \, .
\end{align}
The RPV couplings are therefore proportional to $t_\theta^2$, which is the same parameter that controls the deviation of $m_b/m_\tau$ from 1 at the unification scale.

\section{Phenomenological remarks}
\label{phenoremarks}

The baryon number RPV interactions are subject to stringent low-energy constraints coming mainly from 
proton decay, dinucleon decay, $n$-$\overline{n}$ oscillations, and flavour-violating observables. 
Rescaling the bounds in ref.~\cite{Barbier:2004ez} for superpartners around 500 
GeV and assuming a gravitino heavier than the proton (in order to evade the constraints from proton decay), 
one gets 
\begin{align}
\label{lamsboundsingle1}
\left| \lambda''_{uds} \right| &<  \mathcal{O} (10^{-5} )  \quad [NN \to KK] \, , \\
\label{lamsboundsingle2}
\left| \lambda''_{udb} \right| &<  \mathcal{O} (10^{-3} )  \quad [n-\overline{n}] \, , \\ 
\label{lamsboundsingle3}
\left| \lambda''_{tds} \right| &< \mathcal{O} (10^{-1} ) \quad [n-\overline{n}] \, , \\ 
\label{lamsboundsingle4}
\left| \lambda''_{tdb} \right| &<  \mathcal{O} (10^{-1} ) \quad [n-\overline{n}] \, , 
\end{align}
and 
\begin{align}
\label{lamsboundflavour1}
\left| \lambda''_{cdb} \, \lambda''_{csb}  \right| &< 
\mathcal{O} (10^{-3}) \quad [K-\overline{K}] \, , \\
\label{lamsboundflavour2}
\left| \lambda''_{tdb} \, \lambda''_{tsb} \right| &< 
\mathcal{O} (10^{-3}) \quad [K-\overline{K}] \, , \\
\label{lamsboundflavour3}
\left| \lambda''_{ids} \, \lambda''_{idb}  \right| &< 
\mathcal{O} (10^{-1}) \quad [B^+\rightarrow K^0\pi^+] \, , \\
\label{lamsboundflavour4}
\left| \lambda''_{ids} \, \lambda''_{isb} \right| &< 
\mathcal{O} (10^{-3}) \quad [B^-\rightarrow \phi\pi^-] \, ,
\end{align}
with $i=u,c,t$ for the product of two RPV couplings. 
The bounds quoted above have a strong dependence on the spectrum of the superpartners 
and large uncertainties related to the flavour structure of the soft terms and the hadronic matrix elements. 
However, for the purposes of our discussion, an order of magnitude estimate is sufficient (see e.g.~\cite{Barbier:2004ez} and references therein).

Upper bounds coming from the requirement of not washing out a preexisting baryon asymmetry generated above the electroweak 
scale turn out to give 
$\lambda''< 3 \cdot 10^{-7}$ for sfermion masses of about 1 TeV \cite{Barbier:2004ez}. However, 
such a constraint should be regarded as a sufficient condition for the baryon asymmetry of the Universe not to be erased by RPV interactions, 
rather than a strict bound (see e.g.~the discussion in sect.~4.2.~of \cite{Barbier:2004ez}).

On the other hand, the RPV couplings cannot be too small, if the SUSY searches based on the missing energy signature are to be evaded. This is the case if the NLSP (we assume the LSP to be the gravitino) has a prompt decay 
corresponding to a decay length smaller than about 2 mm \cite{Krnjaic:2013eta}.\footnote{Larger decay lengths give rise to displaced vertices, which require dedicated analysis. 
See, for instance, \cite{Graham:2012th,Allanach:2012vj}.}
This way supersymmetry can be ``hidden'' into QCD backgrounds and the lower bounds on superpartners can be relaxed with respect to the R-parity conserving case. 

To illustrate this point, let us compare the current exclusion limits from LHC in standard MSSM scenarios to the case with bayonic RPV. In the case of the R-parity conserving MSSM, the present lower bounds on the stop and gluino masses are, respectively, $m_{\tilde{t}} \gtrsim 700$ GeV \cite{ATLAS-CONF-2013-024,ATLAS-CONF-2013-037} and $m_{\tilde{g}} \gtrsim 1.3$ TeV \cite{CMS-SUS-13-007,CMS-SUS-12-024}; 
where the bound on the stop is conservative and applies in the case of a light 
neutralino and away from the kinematical configuration $m_{\tilde{t}} \approx m_t + m_{\chi}$. 
In the case of the simplified squark-gluino-neutralino model, one gets $m_{\tilde{g}}, m_{\tilde{t}} \gtrsim 1.5 $ TeV (with only 5.8 fb$^{-1}$ of integrated luminosity and $\sqrt{s} = 8$ TeV) \cite{ATLAS:2012ona}. On the other hand, if we allow the light colored s-particles (gluinos and squarks) to decay promptly via the $u^c d^c d^c$ operator, the bounds are much less stringent. 
For instance, if the stop decays directly into jets neither ATLAS nor CMS can currently place significant limits on the stop 
mass \cite{Evans:2012bf,Aad:2011yh,ATLAS:2012ds,Chatrchyan:2013izb}. The decay of the gluino can proceed either through $\tilde{g} \to \tilde{t} t$ (and consequently $\tilde{t} \to b s$ for example) or directly into jets. In the former case, the bound on the gluino mass is 890 GeV (with 20.7 fb$^{-1}$ and $\sqrt{s} = 8$ TeV) \cite{ATLAS-CONF-2013-007}, while in the latter one, it is 666 GeV (with 4.6 fb$^{-1}$ and $\sqrt{s} = 7$ TeV) \cite{ATLAS:2012dp}.

Let us quantify now the minimal amount of RPV needed in order to have a prompt vertex. 
As a benchmark scenario we consider the case of a right-handed stop NLSP decaying into two SM fermions. 
Though a stop NLSP cannot be achieved in scenarios with universal boundary conditions (see e.g.~\cite{DiazCruz:2007fc}), 
in our case the MSSM soft masses are not SO(10) invariant (even assuming that SUSY is broken above the GUT scale) 
because of the mixed embedding of the MSSM matter fields. 
As shown in \app{stopNLSP}, a right-handed stop NLSP is a realistic possibility. In such a case, the decay length reads 
\begin{equation}
L = 2 \, \textrm{mm} \left( \beta \gamma \right)  \left( \frac{500 \textrm{ GeV}}{m_{\tilde{t}^c}}   \right) \left( \frac{0.9 \cdot 10^{-7}} {\lambda''} \right)^2 \, ,
\end{equation}
where $\beta$ is the velocity of the decaying particle and $\gamma$ is the Lorentz boost factor. 
Hence, a decay length smaller than about 2 mm requires $\lambda'' > \mathcal{O} (10^{-7})$. 

Moreover, it is worth it to mention that the flavour structure of the GUT-induced $\lambda''_{ijk}$ 
which emerges from \eq{lambdasFS}, is of the type
\begin{equation}
\label{structure}
\lambda''_{ijk} \propto \alpha_i \beta_{[j} \gamma_{k]} \, ,
\end{equation}
where $\alpha_i$, $\beta_j$ and $\gamma_k$ are independent 3-vectors 
in the flavour space. This nongeneric structure implies a set of low-energy correlations among the 
RPV couplings. For instance, we find that the following relations
\begin{equation}
\label{corrlambdas}
\frac{\lambda''_{ids}}{\lambda''_{jds}}=\frac{\lambda''_{idb}}{\lambda''_{jdb}}=\frac{\lambda''_{isb}}{\lambda''_{jsb}} \, ,
\end{equation}
must be satisfied for $i,j = u,c,t$. 
Though in principle testable, it would be admittedly difficult (if not impossible) to probe the relations in \eq{corrlambdas} at the LHC,
since one would need to identify the family of the produced quarks.

In the presence of additional assumption on a common origin of the flavour structure of both the SM fermions and the RPV couplings, the RPV couplings also show a hierarchical pattern, as illustrated by the example in \sect{flavormodel}. 
A simple consequence is that a stop will decay predominantly into $\tilde{t} \to b s$. 
 
Finally, let us mention that a hierarchical flavour pattern for $\lambda''$ is also 
predicted in different schemes like minimal flavour violation \cite{Nikolidakis:2007fc,Csaki:2011ge}, partial compositness \cite{KerenZur:2012fr}, 
gauged flavour symmetries \cite{Krnjaic:2012aj,Franceschini:2013ne,Csaki:2013we}, 
and anomalous U(1) symmetries \cite{Florez:2013mxa,Monteux:2013mna}.  
In our case the hierarchical pattern is a direct consequence of the postulated ${\rm SU(3)}_H$ 
symmetry together with the GUT structure of the Yukawa sector. The resulting pattern 
(e.g.~the ratios between the RPV couplings) is then simply different from the existing constructions proposed so far, 
though it is not obvious a priori how to devise phenomenological strategies in order to distinguish them. 

\section{Summary}
\label{conclusions}

Supersymmetric models with R-parity violation have the potential to relieve some of the pressure on the naturalness of supersymmetric extensions of the SM due to the lack of signals at the LHC. This is welcome, as providing a natural framework for electroweak symmetry breaking is one of the main phenomenological motivations of supersymmetry. On the other hand, this requires baryon number violating RPV operators not to be accompanied by lepton number violating ones, which in turn may seem to require giving up another important phenomenological motivation: the possibility to explain the SM fermion gauge quantum numbers within a grand unified framework leading to a successful prediction for the unification of gauge couplings. We have shown that this is not the case. Dimension four lepton number violating interactions can vanish, despite the presence of sizeable baryon number violating interactions and the existence of a grand unified gauge symmetry relating baryon and leptons, in models in which the necessary sources of GUT breaking split the unified multiplets and additional vectorlike matter is added to the MSSM chiral content. 

In particular, we have shown that this can be achieved without fine-tuning or the need of large representations in simple renormalizable SO(10) models in which the adjoint vev is aligned along the 3R or B-L generators. In this context, it is also possible to relate the size of baryonic R-parity violation to the origin of the SM fermion mass hierarchy and to the success (to some extent) of unified relations among third-family fermion masses.

\subsection*{Acknowledgments}

We thank Giorgio Arcadi, Zurab Berezhiani, Roberto Franceschini, Riccardo Torre, Giovanni Villadoro, and Francesco Vissani for useful discussions. The work of L.D.L.\ was supported by the DFG through the SFB/TR 9 ``Computational Particle Physics''. The work of M.N.\ was partly supported by the Danish National Research Foundation under the research grant DNRF:90. The work of A.R.\ was partly supported by the EU Marie Curie ITN ``UNILHC'' (PITN-GA-2009-237920) and the ERC Advanced Grant no. 267985Ê``DaMESyFla''. M.N.\ and A.R.\ thank the Galileo Galilei Institute for Theoretical Physics for the hospitality and the INFN for partial support.

\appendix

\section{Details of the model in \sect{minimalSO10model}} 

\label{MS101616bar}

In this appendix we illustrate the details of the analysis of the model specified by \eq{WRPV1616bar10} and \eq{Wyuk10} of \sect{minimalSO10model}. To identify the light MSSM components populating the SO(10) fields $16_a$, $16$, and $10$, one has to inspect the mass matrices stemming from \eq{WRPV1616bar10} 
upon SO(10)-symmetry breaking. In particular, the piece of superpotential responsible for the nonpure embedding of the MSSM degrees of freedom into the relevant SO(10) representations reads
\begin{multline}
\label{piecesup}
W = \alpha_a \overline{16}\, 45_H 16_a + \beta_a 16_H 16_a 10 \\ + M_{16}\overline{16} 16 + \frac{M_{10}}{2} 10 \, 10 + \ldots \, ,
\end{multline}
where the Higgs superfields $45_H$ and $16_H$ are assumed to pick up a GUT-scale vev 
$\vev{45_H} = V_{45} T_{3R}$ and $\vev{16_H} = V_{16}$ along the ${\rm SU(4)}_{PS} \otimes {\rm SU(2)}_L \otimes {\rm U(1)}_R$ and SU(5) invariant directions, respectively.

The mechanism we are going to consider is based on the fact that the $45_H$ vev picks up the ${\rm SU(2)}_L$ singlet components of $16_a$
and the vev of the $16_H$ picks up the $\overline{5}_{16_a}$ and $5_{10}$ SU(5) components of $16_a$ and $10$. Hence, upon SO(10)-symmetry breaking, \eq{piecesup} leads to the following mass matrices involving the MSSM-like degrees of freedom: 
\begin{align}
\label{Mmdc}
&\begin{pmatrix}
\overline{d}^c_{\overline{16}} Ê& Ê \overline{d}^c_{10} 
\end{pmatrix}
\begin{pmatrix}
V_{45} \, \alpha_a & M_{16} Ê& Ê0  \\
V_{16} \, \beta_a & 0 & M_{10} 
\end{pmatrix}
\begin{pmatrix}
d^c_{16_a} \\ d^c_{16} \\ d^c_{10}
\end{pmatrix} \, , \\
\label{Mml}
&\begin{pmatrix}
\overline{l}_{\overline{16}} Ê& Ê \overline{l}_{10} 
\end{pmatrix}
\begin{pmatrix}
0 & M_{16} Ê& Ê0  \\
V_{16} \, \beta_a & 0 & M_{10} 
\end{pmatrix}
\begin{pmatrix}
l_{16_a} \\ l_{16} \\ l_{10}
\end{pmatrix} \, , 
\\
\label{Mmuc}
&\begin{pmatrix}
\overline{u}^c_{\overline{16}} 
\end{pmatrix}
\begin{pmatrix}
- V_{45} \, \alpha_a & M_{16}  
\end{pmatrix}
\begin{pmatrix}
u^c_{16_a} \\ u^c_{16} 
\end{pmatrix} \, , \\
\label{Mmec}
&\begin{pmatrix}
\overline{e}^c_{\overline{16}} 
\end{pmatrix}
\begin{pmatrix}
V_{45} \, \alpha_a & M_{16}  
\end{pmatrix}
\begin{pmatrix}
e^c_{16_a} \\ e^c_{16} 
\end{pmatrix} \, , \\
\label{Mmq}
&\begin{pmatrix}
\overline{q}_{\overline{16}} 
\end{pmatrix}
\begin{pmatrix}
0 & M_{16}  
\end{pmatrix}
\begin{pmatrix}
q_{16_a} \\ q_{16} 
\end{pmatrix} \, .
\end{align}
Let us leave aside for a while the $d^c$-like states and focus on the others. 
From \eqs{Mml}{Mmq} one can readily find the heavy (GUT-scale) mass eigenstates 
\begin{align}
\label{l5eig}
L_1 & = l_{16}  \, , \\
\label{l4eig}
L_2 & =  \cos\phi \, l_{10} + \sin\phi \, \hat{\beta}_a l_{16_a}  \, , \\[1mm]
\label{uc4eig}
U^c & = \cos\theta \, u^c_{16} - \sin\theta \, \hat{\alpha}_a u^c_{16_a} \, , \\[0.9mm]
\label{ec4eig}
E^c & = \cos\theta \, e^c_{16} + \sin\theta \, \hat{\alpha}_a e^c_{16_a} \, , \\[0.9mm]
\label{q4eig}
Q & = q_{16}  \, ,
\end{align}
where we defined the quantities 
$\tan\theta \equiv V_{45} \alpha / M_{16}$ and $\tan\phi \equiv V_{16} \beta / M_{10}$, 
and the normalized vectors $\hat{\alpha}_a \equiv \alpha_a / \alpha$ and 
$\hat{\beta}_a \equiv \beta_a / \beta$, with $\alpha \equiv \sqrt{\sum_a \alpha_a^2 }$ 
and $\beta \equiv \sqrt{\sum_a \beta_a^2}$.  
The light MSSM components $l_a$, $u^c_a$, $e^c_a$, $q_a$ ($a=1,2,3$), 
can be identified with the linear combinations orthogonal to those in 
\eqs{l5eig}{q4eig}. A possible choice is
\begin{align}
\label{llight}
l_3 &= - \sin\phi \, l_{10} + \cos\phi \, \hat{\beta}_a l_{16_a} \, \\[0.5mm]
\label{uclight}
u^c_3 &= \sin\theta \, u^c_{16} + \cos\theta \, \hat{\alpha}_a u^c_{16_a} \, , \\ 
\label{eclight}
e^c_3 &= - \sin\theta \, e^c_{16} + \cos\theta \, \hat{\alpha}_a e^c_{16_a} \, ,
\end{align}
while the remaining light components: $l_m$, $u^c_m$, $e^c_m$ ($m=1,2$) 
and $q_a$ ($a=1,2,3$) are only contained in the $16_a$.
In particular, we are interested in the projection of the $10$ and $16$ fields on the 
light eigenstates. Inverting the transformations in \eqs{l5eig}{eclight}, we get 
\begin{align}
\label{l10dec}
l_{10} &\rightarrow -\sin\phi \, l_3 \, , \\
\label{l16dec}
l_{16} &\rightarrow 0 \, , \\
\label{uc16dec}
u^c_{16} &\rightarrow \sin\theta \, u^c_3 \, , \\
\label{ec16dec}
e^c_{16} &\rightarrow - \sin\theta \, e^c_3 \, , \\
\label{q16dec}
q_{16} &\rightarrow 0 \, .
\end{align}
The identification of the $d^c$-like light states is more involved. Therefore, let us first consider the simple limit in which the vectors $(\alpha_a)$ and $(\beta_a)$ are orthogonal, before considering the general case. In such a case, the heavy mass eigenstates are 
\begin{align}
\label{dc5eig}
D^c_1 & = \cos\theta \, d^c_{16} + \sin\theta \, \hat{\alpha}_a d^c_{16_a} \, , \\
\label{dc4eig}
D^c_2 & = \cos\phi \, d^c_{10} + \sin\phi \, \hat{\beta}_a d^c_{16_a} \, ,
\end{align}
and the light $d^c_a$ components can be chosen to be
\begin{align}
\label{dclightorth}
d^c_3 &= - \sin\theta \, d^c_{16} + \cos\theta \, \hat{\alpha}_a d^c_{16_a} \, , \\
d^c_2 &= - \sin\phi \, d^c_{10} + \cos\phi \, \hat{\beta}_a d^c_{16_a} \, , 
\end{align} 
while $d^c_1$ is entirely contained in the $16_a$. The projection of the $10$ and $16$ fields on the light $d^c$-like states then reads 
\begin{align}
\label{dc16dec}
d^c_{16} &\rightarrow -\sin\theta \, d^c_3 \, , \\
\label{dc10dec}
d^c_{10} &\rightarrow -\sin\phi \, d^c_2 \, . 
\end{align}
The only renormalizable (RPV) interaction generated by the operator $\lambda \, 16 16 10$ (cf.~\eq{WRPV1616bar10}) is therefore 
\begin{equation}
\label{161610proj}
2 \lambda \sin^2\theta \sin\phi \, u^c_3 d^c_3 d^c_2 \, .
\end{equation}
In the opposite case in which $\alpha_a$ and $\beta_a$ are parallel, both $d^c_{16}$ and $d^c_{10}$ contain only one linear combination of the light fields, and the baryon number violating RPV operator would vanish by antisymmetry. 

Let us now consider the general case. To identify the light $d^c$ eigenstates, it is useful to consider a basis in the SO(10) flavour space in which $\beta_1 = 0$, $\alpha_{1,2} = 0$, so that $(\alpha_a) = (0,0,\alpha_3)$, $\alpha_3 > 0$, $(\beta_a) = (0,\beta_2,\beta_3)$, $\alpha = \alpha_3$, $\beta = (\beta_2^2+\beta_3^2)^{1/2}$. In such a basis, one of the three light eigenstates is $d^c_1$ and the other two are linear combinations of $d^c_{16_2}$, $d^c_{16_3}$, $d^c_{16}$, $d^c_{10}$ orthogonal to the heavy linear combinations (linearly independent but not orthogonal nor normalized)
\begin{align}
\label{eq:heavyD}
D^c_1 &= \alpha V_{45} d^c_{16_3} + M_{16} d^c_{16} \\
D^c_2 &= V_{16}(\beta_3 d^c_{16_3} + \beta_2 d^c_{16_2}) + M_{10} d^c_{10} \, .
\end{align}
A possible choice of the light fields is given by the exterior products
\begin{align}
\label{eq:lightda}
d^c_2 &= (D^c_1  \wedge D^c_2  \wedge d^c_{16_3})/N_2 \\
\label{eq:lightdb}
d^c_3 &= (D^c_1 \wedge D^c_2  \wedge d^c_2)/N_3 \, ,
\end{align}
where $N_2$ and $N_3 $ are normalization factors. The explicit expressions are
\begin{align}
\label{eq:lightd2}
d^c_2 &= \frac{d^c_{16_2}-\hat\beta_2 t_\phi \, d^c_{10}}{(1+(\hat\beta_2 t_\phi)^2)^{1/2}} \\
\label{eq:lightd3}
d^c_3 &= \frac{(1+(\hat\beta_2 t_\phi)^2) (d^c_{16_3}-t_\theta d^c_{16})-\hat\beta_3t_\phi (\hat\beta_2 t_\phi d^c_{16_2} +d^c_{10})}{(1+(\hat\beta_2 t_\phi)^2)^{1/2}(1+t^2_\theta+t^2_\phi+\hat\beta^2_2 t^2_\phi t^2_\theta)^{1/2}} \, ,
\end{align}
from which we get
\begin{multline}
\label{eq:RPVexact}
\lambda 16\, 16\, 10 = \text{heavy} + \\
\frac{2\lambda \hat{\alpha}_{[3}\hat{\beta}_{2]} s_\theta t_\theta t_\phi }{(1+t^2_\theta+t^2_\phi + (1-(\hat\alpha\cdot\hat\beta)^2) t^2_\phi t^2_\theta)^{1/2}} u^c_3 d^c_3 d^c_2.
\end{multline}
The coefficient of the RPV operator in the previous expression is independent of the choice of the two light fields $d^c_3$, $d^c_2$ made in \eqs{eq:lightda}{eq:lightdb}, provided that $d^c_3$, $d^c_2$ are orthonormal and orthogonal to $d^c_{16_1}, D^c_1, D^c_2$. The form in which it is written is independent of the basis in which the vectors $(\alpha_a)$ and $(\beta_a)$ are written, as long as $\alpha_1 = \beta_1 = 0$.

In the $t_\theta^2 \ll 1 \ll t_\phi^2$ limit identified in \sect{flavormodel}, the coefficient of the RPV operator becomes
\begin{equation}
\label{eq:limit}
2\lambda  s_\theta t_\theta \hat{\alpha}_{[3}\hat{\beta}_{2]} \approx 2 \lambda   
\frac{V_{45}^2 \alpha^2}{M^2_{16}} \hat{\alpha}_{[3}\hat{\beta}_{2]} \, .
\end{equation}
We remind that \eq{eq:RPVexact} should be written in terms of the fermion mass eigenstates, which are determined by the SM Yukawas after electroweak symmetry breaking. 

In \sect{minimalSO10model}, we also considered the limit  $M_{16,10} \gg V_{45,16}$. In this limit, corresponding to small angles $\theta$ and $\phi$, the light $d^c_a$ states can be obtained as perturbations of the states $d^c_{16_a}$,
\begin{equation}
\label{dcadeclim}
d^{c}_a \approx d^c_{16_a} 
- \theta \, \hat{\alpha}_a d^c_{16}
- \phi \, \hat{\beta}_a d^c_{10} \, , 
\end{equation}
from which we get
\begin{equation}
\label{161610projdeclim}
\lambda \, 16 16 10 \approx 2 \lambda \theta^2 \phi \, 
\hat{\alpha}_a \hat{\alpha}_b \hat{\beta}_c u^c_a d^c_b d^c_c \, , +\text{heavy}
\end{equation}
which yields the operator $\lambda''_{abc}  u^c_a d^c_b d^c_c$, with 
\begin{equation}
\label{lambdasdeclim}
\lambda''_{abc} = \lambda \theta^2 \phi \, 
\hat{\alpha}_{a} \hat{\alpha}_{[b} \hat{\beta}_{c]} \, ,
\end{equation}
the same expression in \eq{lambdaseffSO10A}, obtained by integrating out the heavy vectorlike fields 
$16 \oplus \overline{16} \oplus 10$ at the SO(10) level. 

\section{Details of the flavor model}
\label{detflavmod}

To obtain \eqs{mtfm}{VCKMfm},
one can follow the procedure illustrated in the previous Appendix with $(\alpha_a) = \alpha (0,0,1)$ and 
$(\beta_a) = \beta (0,\epsilon,1)$ and expand at the leading order in $\epsilon \ll 1$.  In particular, we choose the same basis as in \eqs{eq:lightd2}{eq:lightd3} for the light $d^c$ eigenstates. 
Analogously, the light $l$ eigenstates are defined by replacing $d^c \leftrightarrow l$ and setting $t_\theta = 0$ in \eqs{eq:lightd2}{eq:lightd3}.  
The basis for the other light states follows the conventions given in \app{MS101616bar}.
At the leading order in $\epsilon$, we find
the following projections for the SO(10) current eigenstates onto the light degrees of freedom: 
\begin{itemize}
\item $d^c$-like states
\begin{align}
\label{dclikeproj}
d^c_{16_2} &\rightarrow d^c_2 - \epsilon N c_\theta t^2_\phi d^c_3 \, , \nonumber \\
d^c_{16_3} &\rightarrow N c_\theta d^c_3 \, , \nonumber \\
d^c_{16} &\rightarrow - N s_\theta d^c_3 \, , \nonumber \\
d^c_{10} &\rightarrow  - \epsilon t_\phi d^c_2 -N c_\theta t_\phi d^c_3 \, , 
\end{align}
where $N$ is defined as in \eq{deftttpN3}. 
\item $l$-like states
\begin{align}
\label{llikeproj}
l_{16_2} &\rightarrow l_2 - \epsilon s_\phi t_\phi l_3 \, , \nonumber \\
l_{16_3} &\rightarrow c_\phi l_3 \, , \nonumber \\
l_{16} &\rightarrow 0 \, , \nonumber \\
l_{10} &\rightarrow  - \epsilon t_\phi l_2 - s_\phi l_3 \, . 
\end{align}
Notice that in the $t_\theta \rightarrow 0$ limit, $N \rightarrow c_\phi$.
\item $u^c$-like states
\begin{align}
\label{uclikeproj}
u^c_{16_2} &\rightarrow u^c_2 \, , \nonumber \\
u^c_{16_3} &\rightarrow c_\theta u^c_3 \, , \nonumber \\
u^c_{16} &\rightarrow s_\theta u^c_3 \, . 
\end{align}
\item $e^c$-like states
\begin{align}
\label{eclikeproj}
e^c_{16_2} &\rightarrow e^c_2 \, , \nonumber \\
e^c_{16_3} &\rightarrow c_\theta e^c_3 \, , \nonumber \\
e^c_{16} &\rightarrow - s_\theta e^c_3 \, .
\end{align}
\item $q$-like states
\begin{align}
\label{qlikeproj}
q_{16_2} &\rightarrow q_2 \, , \nonumber \\
q_{16_3} &\rightarrow q_3 \, , \nonumber \\
q_{16} &\rightarrow 0 \, . 
\end{align}
\end{itemize}
By substituting \eqs{dclikeproj}{qlikeproj} into the Yukawa superpotential of \eq{Wyuk10}, we obtain the 
following Yukawa matrices for the second and third families at the leading order in $\epsilon$:
\begin{align}
\label{Yuflavormod}
Y_u &=
\left(
\begin{array}{cc}
\epsilon^2 (2 y_{22}) &  \epsilon (2 c_{\theta} y_{23} +  s_\theta y_2) \\
\epsilon (2 y_{23}) & 2 c_{\theta} y_{33} +  s_\theta y_3 
\end{array}
\right) \, , \\ \nonumber \\
\label{Ydflavormod}
Y_d &= 
\left(
\begin{array}{cc}
\epsilon^2 ( 2y_{22} ) &  \epsilon N  (2 c_\theta y_{23} - s_{\theta} y_2)  \\
\epsilon ( 2 y_{23} ) & N  ( 2 c_\theta y_{33} - s_{\theta} y_3 )
\end{array}
\right) \, , \\ \nonumber \\ 
\label{Yeflavormod}
Y_e &=
\left(
\begin{array}{cc}
\epsilon^2 ( 2y_{22} ) & \epsilon ( 2 c_\theta y_{23} -  s_\theta y_2 )  \\
\epsilon ( 2  c_\phi y_{23} ) & c_\phi ( 2 c_\theta y_{33}  -  s_\theta y_3 )
\end{array}
\right) \, ,
\end{align}
where the basis for $Y_{u,d,e}$ is chosen is such a way that 
the ${\rm SU(2)}_L$ doublets act from the left. 
Notice that $Y_d= Y^T_e$ in the $t_\theta \rightarrow 0$ (and hence $N \rightarrow c_\phi$) limit, 
as expected from the fact that SU(5) is unbroken in this limit. 
Analogously, $Y_d = Y_e$ in the $t_\phi \rightarrow 0$ (and hence $N \rightarrow 1$) limit, 
as expected from the fact the Pati-Salam factor ${\rm SU(4)}_{PS}$ is unbroken in this limit. 

The perturbative diagonalization of \eqs{Yuflavormod}{Yeflavormod} leads to the physical masses and mixings collected in \eqs{mtfm}{VCKMfm} and to  
right-handed rotation matrices for which the ``2--3'' sector has the following form: 
\begin{align}
\label{Vucflavormod}
V_{u^c} & =
\left(
\begin{array}{cc}
1 & \displaystyle- \epsilon \frac{2 y_{23}}{2 c_\theta y_{33} + s_\theta y_3} \\ 
\displaystyle \epsilon \frac{2 y_{23}}{2 c_\theta y_{33} + s_\theta y_3} & 1
\end{array}
\right) \, , \\ \nonumber \\
\label{Vdcflavormod}
V_{d^c} & = 
\left(
\begin{array}{cc}
1 & \displaystyle- \frac{\epsilon}{N} \frac{2 y_{23}}{2 c_\theta y_{33} - s_{\theta} y_3} \\ 
\displaystyle \frac{\epsilon}{N} \frac{2 y_{23}}{2 c_\theta y_{33} - s_{\theta} y_3} & 1
\end{array}
\right) \, .
\end{align}
Finally, the RPV coupling at low energy is obtained by projecting the operator $16\, 16\, 10$ onto the light states:
\begin{equation}
\label{lambdasfmapp}
\lambda \, 16\, 16\, 10 \rightarrow 2 \, \lambda \, \epsilon \, \frac{s_{\theta} t_{\theta} t_{\phi}}{(1 +t_{\theta}^2 + t^2_{\phi})^{1/2} } u^c_3 d^c_3 d^c_2 .
\end{equation}
At the leading order in $\epsilon$, the rotations in \eqs{Vucflavormod}{Vdcflavormod} do not affect the result, which can be obtained expanding  \eq{eq:RPVexact} at the leading order in $\epsilon$. \eqs{lambdatbsfm}{lambdacbsfm} follow.

\section{Stop NLSP}
\label{stopNLSP}

In this appendix we investigate the feasibility of a stop NLSP. For definiteness, 
we consider the model discussed in \sect{flavormodel} with an ${\rm SU(3)}_H$ horizontal 
symmetry broken by two spurions $A$ and $B$.
Let us assume that SUSY is broken at the Planck scale, 
as suggested by the requirement of a gravitino heavier than about 1 GeV for matter stability 
(cf.~\sect{phenoremarks}). 
In such a case, the soft terms are SO(10) invariant and read
\begin{equation}
\label{Lsoft}
\mathcal{L}_{\text{soft}} = - X_{\alpha\beta}  \tilde{16}^*_\alpha \tilde{16}_\beta - Y \, \tilde{10}^* \tilde{10} + \ldots \, ,
\end{equation}
where $16_\alpha = (16_a, 16)$ with $a=1,2,3$ and for simplicity we consider real parameters, 
so that in particular the matrix $X$ is symmetric. 
Because of the ${\rm SU(3)}_H$ flavour symmetry, 
we expect the following texture 
for the coefficients $X_{\alpha\beta}$ and $Y$ (recall that $16$ and $10$ are invariant under 
${\rm SU(3)}_H$, while $16_a$ transforms as a triplet): 
\begin{equation}
\label{XYsoft}
X = \tilde{m}^2
\left(
\begin{array}{cccc}
a & 0 & 0 & 0 \\
\cdot & a & \epsilon & \epsilon  \\ 
\cdot & \cdot & b & c \\  
\cdot & \cdot & \cdot & d 
\end{array}
\right) \, , \qquad Y = \mathcal{O}(\tilde{m}^2) \, ,
\end{equation}
where $a,b,c,d$ are $\mathcal{O}(1)$ numbers, $\epsilon \approx \sqrt{m_c/m_t} \approx 0.06$ (cf.~\eqs{mtfm}{mcfm}) 
and, consistently with the analysis in \sect{flavormodel}, we neglected the small 
mixing with the first family. 
Notice that the flavour symmetry provides an organization of the flavour structure of the soft terms 
and hence a protection from otherwise anarchical gravity-induced FCNC. 

To identify the soft masses relative to the MSSM fermion superpartners, we 
project \eq{Lsoft} onto the three light eigenstates, according to the expressions in \eqs{dclikeproj}{qlikeproj} of \app{detflavmod}. 
Since we are only interested in the eigenvalues of the sfermion mass matrices, 
we expand at the zeroth order in $\epsilon$, thus obtaining  
\begin{align}
\label{msoftq}
\tilde{m}^2_q & \approx
\text{diag}
\left(
X_{11}, X_{22}, X_{33} 
\right) \, , \\
\label{msoftec}
\tilde{m}^2_{e^c} &\approx
\text{diag}
\left(
X_{11}, X_{22}, c^2_\theta X_{33} + s^2_\theta X_{44} - s_{2\theta} X_{34} 
\right) \, , \\
\label{msoftuc}
\tilde{m}^2_{u^c} & \approx
\text{diag}
\left(
X_{11}, X_{22}, c^2_\theta X_{33} + s^2_\theta X_{44} + s_{2\theta} X_{34} 
\right) \, , \\
\label{msoftl}
\tilde{m}^2_{l} & \approx
\text{diag}
\left(X_{11}, X_{22}, c^2_\phi X_{33} + s^2_\phi Y
\right) \, ,
\end{align}
\begin{widetext}
\begin{equation}
\label{msoftdc}
\tilde{m}^2_{d^c} \approx
\text{diag}
\left( 
X_{11}, X_{22}, N^2 ( c^2_\theta X_{33} + s^2_\theta X_{44} - s_{2\theta} X_{34} + c^2_\theta t_\phi^2 Y) 
\right) \, ,
\end{equation}
\end{widetext}
where the parameters $\theta$, $\phi$, and $N$ were defined in \eq{deftttpN3}.  

Because of the renormalization group effects, the lightest sfermion is a third-family sfermion in a wide region of the parameter space 
(depending on the relative size of $X_{11}, X_{22}, X_{33}$). Let us then focus on the third-family soft masses.
Notice, also, that in the SO(10)-preserved limit ($\theta = \phi = 0$) the boundary conditions for the third-generation sfermion masses are universal. 
Working in the phenomenological limit $t^2_\theta \ll 1 \ll t^2_\phi$ identified in \sect{flavormodel}, they read explicitly 
\begin{align}
\label{q3ms}
\tilde{m}^2_{q_3} &\approx X_{33} \, , \\
\label{taucms}
\tilde{m}^2_{\tau^c} &\approx X_{33} - 2 \theta X_{34}  \, , \\
\label{tcms}
\tilde{m}^2_{t^c} &\approx X_{33} + 2 \theta X_{34}  \, , \\
\label{l3ms}
\tilde{m}^2_{l_3} &\approx Y \, , \\
\label{bcms}
\tilde{m}^2_{b^c} &\approx Y \, .
\end{align}
In the limit of $\theta = 0$, one predicts $b$-$\tau$ unification at the GUT scale (cf.~the discussion in \sect{flavormodel}). 
However, $\theta$ does not need to be exactly zero since departures from $b$-$\tau$ unification at the GUT scale 
are in general possible. 
Assuming $\theta X_{34} < 0$ and $Y$ larger than $X_{33}$, 
a cancellation at the level of $50 \%$ in the expression for $\tilde{m}^2_{t^c}$ in \eq{tcms} is sufficient in order to achieve a stop NLSP, 
even after taking into account the running of the soft masses from the GUT to the soft scale.

\bibliography{bibliography}
\bibliographystyle{h-physrev5}

\end{document}